\documentclass[aps,journal=prb,twocolumn]{revtex4-1}
\usepackage{graphicx}
\usepackage{epstopdf}
\usepackage{tabularx}
\usepackage{graphicx}

\begin{document}
\title{Origin of Negative Longitudinal Piezoelectric Effect}

\author{Shi Liu}
\email{sliu@carnegiescience.edu}
\affiliation{Extreme Materials Initiative, Geophysical Laboratory, Carnegie Institution for Science, Washington, D.C. 20015-1305 USA}
\author{R. E. Cohen}
\email{rcohen@carnegiescience.edu}
\affiliation{Extreme Materials Initiative, Geophysical Laboratory, Carnegie Institution for Science, Washington, D.C. 20015-1305 USA}
\affiliation{Department of Earth- and Environmental Sciences, Ludwig Maximilians Universit\"{a}t, Munich 80333, Germany}

\date{\today}
\begin{abstract}{
Piezoelectrics with negative longitudinal piezoelectric coefficients will contract in the direction of an applied electric field. Such piezoelectrics are thought to be rare but there is no fundamental physics preventing the realization of negative longitudinal piezoelectric effect in a single-phase material. Using first-principles calculations, we demonstrate that several hexagonal $ABC$ ferroelectrics possess significant negative longitudinal piezoelectric effect. The data mining of a first-principles-based database of piezoelectrics reveals that this effect is a general phenomenon. The origin of this unusual piezoelectric response relies on the strong ionic bonds associated with small effective charges and rigid potential energy surfaces. Moreover, ferroelectrics with negative longitudinal piezoelectric coefficients show anomalous pressure-enhanced ferroelectricity. Our results offer design principles to aid the search of new piezoelectrics for novel electromechanical device applications. 
}
\end{abstract}
\maketitle

Piezoelectrics are a class of functional materials that can convert electrical energy to mechanical energy and vice versa. They serve as critical components in many modern devices, ranging from medical ultrasound, fuel injectors, and SONAR, to vibration-powered electronics~\cite{Mao14p1301624}. Piezoelectricity is usually gauged by the piezoelectric coefficient that characterizes how the polarization changes in response to a stress or strain. This leads to two piezoelectric equations: one links the induced polarization in direction $\alpha$ ($\Delta P_{\alpha}$) with an applied stress with component $j$ ($\sigma_j$ in Voig notion), described by $\Delta P_{\alpha}$ = $d_{\alpha j}\sigma_j$, where $d_{\alpha j}$ is the piezoelectric strain coefficient; another one connects $\Delta P_{\alpha}$  with strain ($\eta$), given by $\Delta P_{\alpha} = e_{\alpha j}\eta_j$, where $e_{\alpha j}$ is the piezoelectric stress coefficient~\cite{Nye1985Book, Bellaiche02p19}. Both $d_{\alpha j}$ and $e_{\alpha j}$ are third-rank tensors, and they are related to each other via elastic compliances $d_{\alpha j}$ = $S_{jk}e_{\alpha k}$. 

Piezoelectricity is a bulk effect~\cite{Martin72p1607} and contains a clamped-ion contribution evaluated at vanishing microscopic strain and an internal-strain contribution due to the relative displacement of sublattices in response to the macroscopic strain~\cite{deGironcoli89p2853,SghiSzab98p4321}. The ``improper'' piezoelectric tensor, defined as ${\partial P_{\alpha}}/{\partial \eta_{j}}$, includes the contributions to the linear order due to polarization rotation or dilation~\cite{Nelson76p1785,SghiSzab98p4321}.  The modern theory of polarization ~\cite{KingSmith93p1651,KingSmith94p5828,Resta93p133,Resta94p899} shows that polarization is only well defined modulo a polarization quantum, and takes on a lattice of values ($P^{b}$, where $b$ is a ``branch" label). The ``proper" piezoelectric coefficient that measures the adiabatic change of the current density resulting from a slow crystal deformation is well defined and does not suffer from any branch dependence.  It is the ``proper" piezoelectric coefficient that should be compared to experiments where piezoelectric coefficients are measured based on the charge flow through the sample~\cite{Vanderbilt00p147}.  In this work, we focus on proper piezoelectric coefficients unless explicitly stated otherwise.

The longitudinal piezoelectric coefficients ($e_{33}$ and $d_{33}$, assuming the polar axis is in the $z$ direction) are almost always positive:  consequently, a tensile strain (uniaxial stretching) increases the polarization, or equivalently, the lattice expands along the direction of an applied electric field, as one would expect from the displacements of charged ions in electric fields.  One well-known exception is ferroelectric polymer poly(vinylidene fluoride) (PVDF) and its copolymers which possess {\em negative} longitudinal piezoelectric effect (NLPE): the polymers will contract in the direction of an applied electric field~\cite{Furukawa84p829,Bystrov13p3591,Katsouras15p78}.  The NLPE in PVDF is attributed to the unique microstructures of PVDF polymers with intermixed crystalline lamellae and amporphous regions~\cite{Katsouras15p78}. Though counterintuitive, the NLPE has received very little attention. In fact, no measurements of NLPE for single-phase materials have been reported in the literature. Previous first-principles studies reported that several III-V zinc-blende compounds ({\em e.g.}, GaAs and GaSb) have small negative $e_{33}$ ($\approx-0.1$ C/m$^2$, where the threefold axis of the cubic zinc-blende unit cell is assumed as the $z$ direction)~\cite{Bernardini97p10024}; wurtzite BN has a relatively large (more negative) $e_{33}$ = $-0.94$ C/m$^2$~\cite{Shimada98p1421,Shimada06p358}. However, the origin of NLPE in single-phase materials is not clear. \\

\begin{table*}
\centering
\caption{Theoretical values of piezoelectric coefficients $e_{33}$ (C/m$^2$) and $d_{33}$ (pC/N), zero-stress dielectric constant $\varepsilon_{33}$, and electromechanical coupling coefficient $k_{33}$ for $ABC$ ferroelectrics. Materials with negative $e_{33}$ have the labels underscored. }
\begin{tabularx}{.8\textwidth}{XXXXXXX}
\hline 
\hline
  $ABC$&  $e_{33}$ & $d_{33}$ & $S_{33}$ &$\varepsilon_{33}$ &  $k_{33}$ \\
  \hline
 $\underline{\rm LiBeP} $ & -0.49 & -3.42 & 0.58 & 16.6  & 0.117\\
LiMgP  & 1.51 & 24.66 & 1.64 & 20.6  & 0.451\\
LiZnP  & 0.57 & 3.44 & 0.70 & 24.3  & 0.088\\
$\underline{\rm LiBeAs}$  & -0.48 & -3.70 & 0.65 & 17.6  & 0.117\\
LiMgAs  & 1.54 & 28.60 & 1.88 & 23.2  & 0.461\\
LiZnAs  & 0.37 & 2.25 & 0.81 & 28.0  & 0.050\\
$\underline{\rm LiBeSb}$  & -0.66 & -5.31 & 0.74 & 19.0  & 0.150\\
$\underline{\rm LiZnSb}$  & -0.56 & -5.76 & 0.92 & 30.6  & 0.115\\
$\underline{\rm LiBeBi}$  & -0.76 & -7.26 & 0.89 & 23.0  & 0.171\\
NaMgP  & 0.58 & 7.80 & 1.83 & 20.6  & 0.135\\
NaMgAs  & 0.57 & 8.44 & 1.96 & 21.5  & 0.138\\
NaMgSb  & 0.49 & 8.09 & 2.06 & 20.8  & 0.131\\
$\underline{\rm NaZnSb}$  & -1.04 & -17.30 & 1.43 & 25.0  & 0.308\\
$\underline{\rm KMgSb}$  & -0.42 & -19.15 & 3.14 & 19.6  & 0.259\\
\hline
\hline
\end{tabularx} 
\end{table*}

We first consider the piezoelectric properties of a family of $ABC$ ferroelectrics recently discovered through first-principles high-throughput density functional theory (DFT) computations~\cite{Bennett12p167602}. As a variant of the half-Heusler structure, hexagonal $ABC$ ferroelectrics are in the polar space group  $P6_3mc$, same as that of wurtzite BN, and the unit cell has six atoms with $B$ and $C$ atoms forming buckling honeycomb layers separated by layers of  $A$ atoms. Several $ABC$ ferroelectrics ({\em e.g.}, LiBeSb) are found to possess the so-called ``hyperferroelectricity" characterized by the persistent polarization even at the zero displacement field boundary condition~\cite{Garrity14p127601,Fu14p164104,Liu17p24403}.  It was suggested that the short-range repulsion is likely to play an important role in driving the ferroelectric instability in the paraelectric $P6_3/mcm$ phase~\cite{Li16p34085}.  

Our DFT calculations reveal that a few $ABC$ ferroelectrics possess significant NLPE. Among all studied $ABC$ ferroelectrics, KMgSb has the most negative $d_{33}$ of $-19$ pC/N, whereas LiMgAs has the most positive $d_{33}$ of 29 pC/N, both compared well to known piezoelectrics such as ZnO with $d_{33} \approx 20$ pC/N~\cite{Catti03p2183,Madelung93Book,Coleman06p1}. By screening through a database of calculated intrinsic piezoelectric constants for 941 inorganic crystalline compounds~\cite{deJong15p150053}, we find that the NLPE is a general phenomenon as more than 90 compounds have negative $e_{33}$ and 52 of them were already reported in the {\em Inorganic Crystal Structural Database} (ICSD). Moreover, we find that the electric polarization of several $ABC$ ferroelectrics increases in magnitude as a hydrostatic pressure is applied, making them appealing for high-pressure applications. \\

 All first-principles calculations are carried out using ABINIT~\cite{Gonze02p478,Gonze09p2582} with local density approximation and an $8\times8\times8$ Monkhorst-Pack sampling for the hexagonal lattice. We used ultrasoft pseudopotentials from the Garrity, Bennett, Rabe, Vanderbilt (GBRV) high-throughput pseudopotential set~\cite{Garrity14p446} and a plane-wave cutoff of 25 Ha and a charge density cutoff of 125 Ha. A force convergence threshold of 1.0$\times 10^{-5}$  Ha/Bohr and Gaussian smearing of 1 mHa are used to fully relax the lattice constants and atomic positions. The hexagonal crystal, with the $c$ axis in the $z$ direction, has three independent piezoelectric coefficients, $e_{15}$, $e_{31}$, and $e_{33}$. We focus on the longitudinal piezoelectric coefficient $e_{33}$ and $d_{33}$, which are evaluated with density functional perturbation theory (DFPT)~\cite{Gonze97p10355} and a larger plane-wave cutoff of 30 Ha and charge density cutoff of 250 Ha. The electromechanical coupling factor $k_{\alpha j}$,  an important figure of merit for piezoelectrics that measures the effectiveness of energy conversion, can be estimated with $k_{33}={|d_{33}|} /{\sqrt{\varepsilon_{33} S_{33}}}$, where $\varepsilon$ is the zero-stress dielectric tensor and $S$ is the compliance tensor~\cite{Wu05p035105}. \\

\begin{table*}
\centering
\caption{Born effective charges $Z_{33}^*$, clamped-ion $\bar{e}_{33}$ (C/m$^2$), internal-strain $e_{33}'$ (C/m$^2$), total $e_{33}$ (C/m$^2$) for $ABC$ ferroelectrics. Materials with negative $e_{33}$ have the labels underscored.}
\begin{tabularx}{.8\textwidth}{XXXXXXXXX}
\hline \hline
  $ABC$& $Z_{33}^*(A)$  & $Z_{33}^*(B)$ & $Z_{33}^*(C)$ &  $\frac{\partial u_3(B)}{\partial \eta_3}$ &  $\frac{\partial u_3(C)}{\partial \eta_3}$ &$\bar{e}_{33}$ & $e_{33}'$ & $e_{33}$ \\
  \hline
 $\underline{\rm LiBeP} $ &  1.18 & 1.04 & -2.22 & -0.11 & -0.07 & -0.63 & 0.14 & -0.49\\
LiMgP &  0.93 & 1.85 & -2.77 & -0.19 & -0.40 & -0.11 & 1.62 & 1.51\\
LiZnP &  1.22 & 1.58 & -2.80 & -0.11 & -0.22 & -0.62 & 1.19 & 0.57\\
$\underline{\rm LiBeAs}$ &  1.21 & 0.99 & -2.20 & -0.09 & -0.07 & -0.65 & 0.17 & -0.48\\
LiMgAs &  0.94 & 1.84 & -2.78 & -0.21 & -0.43 & -0.12 & 1.66 & 1.54\\
LiZnAs &  1.29 & 1.66 & -2.95 & -0.13 & -0.24 & -0.73 & 1.10 & 0.37\\
$\underline{\rm LiBeSb}$ &  1.30 & 0.55 & -1.85 & 0.02 & -0.01 & -0.70 & 0.04 & -0.66\\
$\underline{\rm LiZnSb}$ &  1.43 & 1.39 & -2.82 & 0.06 & -0.04 & -0.95 & 0.39 & -0.56\\
$\underline{\rm LiBeBi}$ &  1.35 & 0.61 & -1.96 & 0.04 & 0.01 & -0.76 & 0.01 & -0.76\\
NaMgP &  0.99 & 1.92 & -2.90 & -0.08 & -0.19 & -0.24 & 0.81 & 0.58\\
NaMgAs &  1.01 & 1.90 & -2.91 & -0.09 & -0.21 & -0.26 & 0.83 & 0.57\\
NaMgSb &  1.10 & 1.85 & -2.95 & -0.04 & -0.19 & -0.32 & 0.82 & 0.49\\
$\underline{\rm NaZnSb}$  &  1.06 & 1.76 & -2.81 & -0.04 & -0.01 & -0.95 & -0.09 & -1.04\\
$\underline{\rm KMgSb}$ &  0.70 & 2.01 & -2.71 & -0.10 & -0.05 & -0.39 & -0.03 & -0.42\\
\hline
PbTiO$_3$&3.53  & 5.51 & -4.61 &0.22 & -0.31& -0.86 & 6.06 & 5.20\\
\hline
\hline
\end{tabularx} 
\end{table*}

\begin{figure}[b]
\centering
\includegraphics[scale=0.5]{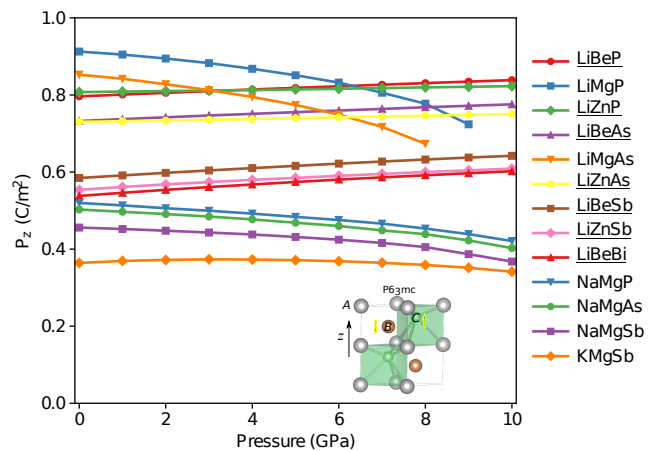}\\
 \caption{ Polarization along the $c$ axis ($P_z$) of $ABC$ ferroelectrics as a function of hydrostatic pressure. Materials exhibiting pressure-enhanced ferroelectricity have the labels underscored. The electrical polarization of KMgSb increases with the pressure until 3 GPa. The inset shows the structure of hexagonal $ABC$ ferroelectrics.}
  \label{PolvsP}
 \end{figure}

We first study 14 $ABC$ semiconducting ferroelectrics~\cite{Bennett12p167602} with DFPT and calculate the piezoelectric constants, compliances, stress-free dielectric constants, and electromechanical coupling factors, reported in Table I. The values of $|e_{33}|$ range from 0.37 to 1.54 C/m$^2$, and $|d_{33}|$ range from 2.25 to 28.6 pC/N, compared well with many III-V nitrides such as AlN, GaN, and InN ($e_{33}$ $\approx$ $1.0-1.5$ C/m$^2$ and $d_{33}$  $\approx$ $2-8$ pC/N)~\cite{Bernardini97p10024,Bernardini02p4145,Shimada06p358}. Notably, the piezoelectric properties of LiMgP ($e_{33}$ = 1.5 C/m$^2$, $d_{33}$ = 25 pC/N, $k_{33}$ = 0.45) compare favorably with classic piezoelectric ZnO ($e_{33}$ = 0.96 C/m$^2$, $d_{33}$ = 12.3 pC/N, $k_{33}$ = 0.41)~\cite{Wu05p035105}, and potentially can serve as a new lead-free high-performance piezoelectric made of earth-abundant elements. Moreover, a few $ABC$ ferroelectrics exhibit NLPE characterized by negative $e_{33}$ and $d_{33}$, and NaZnSb has the most negative $e_{33}$ of $-1.04$ C/m$^2$ and KMgSb has the most negative $d_{33}$ of $-19$ pC/N. 

To understand the origin of NLPE, we decompose $e_{33}$ into two terms~\cite{SghiSzab98p4321}, 
\begin{equation}
e_{33} = \bar{e}_{33} + e_{33}'
\end{equation}
where $\bar{e}_{33}$ is the clamped-ion term computed with the internal atomic coordinates ($u$) fixed at their zero-strain values and 
\begin{equation}
e_{33}' = \sum_s \frac{ec}{\Omega}Z_{33}^*\frac{\partial u_3(s)}{\partial \eta_3}
\end{equation}
is the internal-strain term arising from the internal microscopic atomic relaxations in response to a macroscopic strain $\eta_3$ applied in the $z$ direction. Here $s$ runs over the atoms in the unit cell of volume $\Omega$, $e$ is the electron charge, $c$ is the lattice constant along the polar axis of the hexagonal unit cell, and $Z_{33}^*$ is the Born effective charge associated with the displacement of $u_3$. We find that for $ABC$ ferroelectrics with NLPE, the negative clamped-ion $\bar{e}_{33}$ dominates the total response, whereas the internal-strain contribution $e_{33}'$, though being positive for most compounds, is not large enough to compensate the negative $\bar{e}_{33}$ (Table II). From equation (2), the value of $e_{33}'$ depends on the values of effective charges (``dielectric effect") and the coupling of internal coordinates with the macroscopic strain (``elastic effect")~\cite{Iniguez03p224107}. One can see that the Born effective charges of $ABC$ ferroelectrics (Table II) are close to the nominal ionic charges, in sharp contrast with typical ferroelectric perovskites such as PbTiO$_3$, where the Born effective charges are much larger than the formal charges. This suggests that the ferroelectricity in hexagonal $ABC$ is most likely driven by the geometric ionic size effect~\cite{VanAken04p164,Ederer04p849,Tohei09p144125,Garcia-Castro14p104107} rather than the chemical bonding effect such as the $p$-$d$ orbital hybridization in transition metal oxides~\cite{Cohen92p136}. Moreover, the response of internal coordinates to the macroscopic strain characterized by ${\partial u_3(s)}/{\partial \eta_3}$ is also substantially smaller than those in PbTiO$_3$ ({\em e.g.}, ${\partial u_3({\rm Pb})}/{\partial \eta_3}$ = 0.36 and ${\partial u_3({\rm Ti})}/{\partial \eta_3}$ = 0.22). Taking LiBeSb as an example, the total response $e_{33} = -0.66$ C/m$^2$ is mainly associated with $\bar{e}_{33} = -0.7$ C/m$^2$, and the small $e_{33}'$ is due to the small effective charges,  $Z_{33}^*({\rm Be}) = 0.55~e$ and $Z_{33}^*({\rm Sb}) = -1.85~e$,  and nearly zero ${\partial u_3(s)}/{\partial \eta_3}$. 
We also note that for $ABC$ ferroelectrics with positive $e_{33}$, the piezoelectric response is mainly due to the displacement of $C$ atoms, as reflected by the relatively large $Z_{33}^*(C)$ and ${\partial u_3(C)}/{\partial \eta_3}$.\\

\begin{figure}[t]
\centering
\includegraphics[scale=0.8]{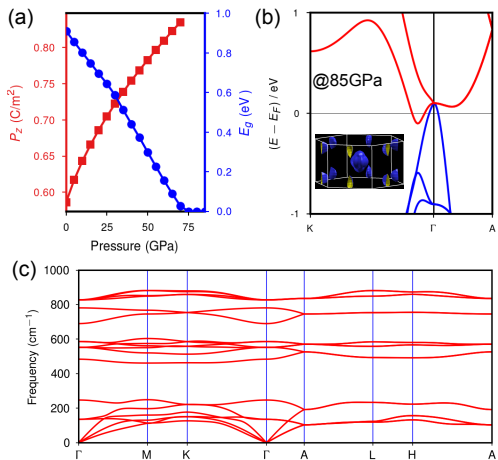}\\
 \caption{(a) Ferroelectric polarization and band gap of LiBeSb as a function of hydrostatic pressure. (b) Electronic band structure and (c) phonon spectrum of LiBeSb in the polar space group $P6_3mc$ at 85 GPa. The inset shows the Fermi surface near the $\Gamma$ point.}
  \label{LiBeSb}

 \end{figure}
 
Our first-principles calculations indicate that the NLPE occurs when the negative clamped-ion piezoelectric response dominates over the internal-strain contribution. It is therefore more likely to realize NLPE in piezoelectrics with strong ionic bonds which may be associated with small effective charges and small atomic relaxations in response to a macroscopic strain. Utilizing a database of piezoelectrics recently developed via high-throughput DFT calculations~\cite{deJong15p150053}, we search for materials with negative $e_{33}$ through a large pool of compounds (941 inorganic compounds). We focus on materials with $|e_{33}|$ being the largest tensor element and $e_{33} < -0.1$ C/m$^2$. Much to our surprise, we find 93 compounds have  negative $e_{33}$, among which 52 compounds were documented in ICSD (see supplementary materials). Notable examples are AgNO$_3$ (ICSD code 374, $R3c$, $e_{33} = -0.60$ C/m$^2$), VPO$_3$ (ICSD code 152278, $P4mm$, $e_{33} = -2.05$ C/m$^2$), and KIO$_3$ (ICSD code 97995, $R3m$, $e_{33} = -1.36$ C/m$^2$).\\

One feature of ferroelectrics with NLPE is the increase of electrical polarization with the decrease of lattice constant along the polar axis, which hints toward the presence of unusual pressure dependence of ferroelectricity. As the hydrostatic pressure generally reduces the lattice constants, ferroelectrics with NLPE may have the polarization enhanced at high pressure.  This is different from conventional ferroelectrics in which the polarization will decrease and eventually vanish with increasing hydrostatic pressure~\cite{Samara71p277,Samara75p1767,Kornev05p196804,Wu05p037601}. 

We calculate the electrical polarization as a function of hydrostatic pressure for 13 $ABC$ ferroelectrics that remain insulating between 0-10 GPa (Fig~\ref{PolvsP}).  As expected, LiBeP, LiBeAs,  LiBeSb, LiZnSb, LiBeBi, and KMgSb, all possessing NLPE,  have the polarization enhanced with the increasing pressure. There are two exceptions: LiZnP and LiZnAs have positive $e_{33}$ while they also show pressure-enhanced ferroelectricity. This can be understood by $\Delta P_3 = 2e_{31}^{\rm IP}\eta_1 + e_{33}^{\rm IP}\eta_3$, where $e_{31}^{\rm IP}$ and  $e_{33}^{\rm IP}$ are the improper piezoelectric constants and $\eta_1$ ($\eta_1 = \eta_2$) and $\eta_3$ are pressure-induced strains. Here the improper piezoelectric coefficients are used because they include the pure volume effect on the polarization. The proper to improper piezoelectric coefficients are related by the spontaneous polarization, $e_{31}^{\rm IP} = e_{31} - P_3$ and $e_{33}^{\rm IP} = e_{33}$~\cite{Vanderbilt00p147}. Therefore, the pressure-enhanced polarization in LiZnP and LiZnAs is due to the negative $e_{31}^{\rm IP}$ = $-0.47$ C/m$^2$ for LiZnP and $e_{31}^{\rm IP}$ = $-0.38$ C/m$^2$ for LiZnAs. We note that similar pressure-enhanced ferroelectricity is reported for improper ferroelectrics such as hexagonal $R$FeO$_3$ ($R$ = Ce, Gd, Lu), where a zone-boundary phonon mode associated with the tilt of FeO$_5$ bipyramid is responsible for the anomalous pressure dependence~\cite{Xu14p205122}. However, $ABC$ ferroelectrics are proper ferroelectrics and the zone-center $\Gamma_2^-$ mode is the primary order parameter responsible for the  transition from paraelectric $P6_3/mmc$ phase to polar $P6_3mc$ phase.

Further investigations of $ABC$ ferroelectrics at higher pressure up to 85 GPa reveal that the ferroelectric polarization can be extremely robust. Taking LiBeSb as an example, we find that the polarization keeps increasing with increasing pressure (Fig.~\ref{LiBeSb}a). By the meanwhile, the band gap becomes smaller and eventually closes at 75 GPa. Electronic band structure and phonon calculations confirm that LiBeSb at 85 GPa is a stable semi-metal with bands across the Fermi energy, though the structure remains in the polar space group $P6_3mc$ (Fig.~\ref{LiBeSb}b and c). This is surprising because the long-range electrostatic dipole-dipole interaction that favors the break of inversion symmetry should be screened out by the free electrons in metal. The observation that  LiBeSb at 85 GPa is a non-centrosymmetric semi-metal further suggests that the ferroelectricity is likely due to geometric effect which is not coupled with states near the Fermi level. \\

In summary, we have explained the origin of negative longitudinal piezoelectric effect with first-principles density functional theory calculations. This counterintuitive piezoelectric response turns out to be a general phenomenon arising from the negative clamped-ion piezoelectric response. The data mining of a database of piezoelectrics leads to more than 90 compounds possessing negative $e_{33}$. The  electrical polarization of $ABC$ ferroelectrics is robust against pressure, and the pressure-enhanced ferroelectricity is intimately related to their unusual piezoelectric properties.  We hope that this work will inspire future experimental studies on negative longitudinal piezoelectric effect, which may offer novel avenues for designing nanoscale electromechanical devices. \\

This work is partly supported by US Office of Naval Research Grants N00014-12-1-1038 and N00014-14-1-0561. 
SL and REC are supported by the Carnegie Institution for Science. REC is also supported by the
European Research Council Advanced Grant ToMCaT. Computational support was provided by theß US DOD through a Challenge Grant from the HPCMO. 
%

\newpage
 
\end{document}